\documentclass[twoside,twocolumn,aps,showpacs, showkeys]{revtex4}
\usepackage{graphicx}
\usepackage{amsmath, amsthm, amsfonts,amssymb}
\usepackage{graphicx}

\begin{document}
\title[]{Long-term Memory and Volatility Clustering in Daily and High-frequency Price Changes}

\author{GabJin \surname{Oh}}
\email{gq478051@postech.ac.kr} \affiliation{Asia Pacific Center for Theoretical Physics \& NCSL, Department of Physics, Pohang University of Science
and Technology, Pohang, Gyeongbuk, 790-784, Korea}

\author{Seunghwan \surname{Kim}}
\email{swan@postech.ac.kr} \affiliation{Asia Pacific Center for Theoretical Physics \& NCSL, Department of Physics, Pohang University of Science
and Technology, Pohang, Gyeongbuk, 790-784, Korea}

\author{Cheol-Jun \surname{Eom}}
 \email{shunter@pusan.ac.kr}\affiliation{Division of Business Administration, Pusan National University, Busan 609-735, Korea}

\received{06 06 2006}

\begin{abstract}
We study the long-term memory in diverse stock market indices and foreign exchange rates using the Detrended Fluctuation Analysis(DFA). 
For all daily and high-frequency market data studied, no significant long-term memory property is detected in the return series, 
while a strong long-term memory property is found in the volatility time series. 
The possible causes of the long-term memory property are investigated using the return data filtered by the AR(1) model, reflecting the short-term memory property, 
and the GARCH(1,1) model, reflecting the volatility clustering property, respectively. 
Notably, we found that the memory effect in the AR(1) filtered return and volatility time series remains unchanged, 
while the long-term memory property either disappeared or diminished significantly in the volatility series of the GARCH(1,1) filtered data. 
We also found that in the high-frequency data the long-term memory property may be generated by the volatility clustering as well as higher autocorrelation. 
Our results imply that the long-term memory property of the volatility time series can be attributed to the volatility clustering observed in the financial time series.
\\



\end{abstract}

\maketitle

\section{INTRODUCTION}

Recently, studies on the hidden or less understood features of financial time series such as the long-term memory 
have been attempted through an interdisciplinary approach with much attention [1,2]. 

The presence of the long-term memory means that the market does not immediately 
respond to an amount of information flowing into the financial market, 
but reacts to it gradually over a period. Therefore, past price changes can be used as a significant information for the 
prediction of future price changes. 
In addition, this observation is significant in that it can provide negative evidence 
as well as a new perspective to the Efficient Market Hypothesis (EMH) [6]. 
Previous works have not found a significant long-term memory in the returns of financial time series. 
However, most of their volatility time series are found to show a strong long-term memory [13-21]. 
The analysis of the return time interval in the stock returns showed that the long-term memory of the volatility is closely related to the 
clustering effect of return time intervals [11,12].
In order to investigate the presence of the long-term memory property in financial time series and its possible causes, we used diverse daily 
and high-frequency market indices. We used daily data from eight international stock market indices from 1991 to 2005 
and the foreign exchange rates of 31 countries to the US dollar from 1990 to 2000, 
as well as high-frequency data including the KOSPI 1-minute market index from 1995 to 2002, 
the KOSDAQ 1-minute market index from 1997 to 2004, and the foreign exchange rates of six nations 
to the US dollar from 1995 to 2004. To quantify the long-term memory property in financial time series, we utilized the Detrended Fluctuation Analysis (DFA), 
which was introduced by Peng {\em et al.} [22].
Previous studies on financial time series has found diverse features that deviate from the random walk process, 
such as autocorrelation, volatility clustering, fat tails and so forth, which were called as stylized facts [3,4,5]. 

Among the several stylized facts, we used data filtered through the Autoregressive (AR) model and the Generalized Autoregressive Conditional heteroscedasticity 
(GARCH) model, reflecting short-term memory and volatility clustering, respectively, which are widely used 
in the financial field [8,9].   
We found that for all types of data used in this paper, Hurst exponents of the return series 
follow a random walk process, while the volatility series exhibit the long memory property with 
Hurst exponents in $0.7 \leq H \leq 0.9$. In order to test the possible causes of the occurrence of the long-term memory property observed in the volatility time series, 
we used the returns series filtered through AR(1) and GARCH(1,1) models.
The long-term memory property was no longer observed in the volatility time series of the daily data 
filtered through the GARCH(1,1) model, while it is considerably reduced with $0.6 \leq H \leq 0.65$ in the volatility time series of the 
high-frequency GARCH(1,1) filtered data. On the other hand, 
the volatility time series of the daily and high-frequency data filtered through the AR(1) model 
still show the strong long-term memory property observed in the volatility time series of 
original time series. 
In order to investigate the possible causes of the differences between results from the daily and high-frequency data, 
we studied the autocorrelation function for the KOSPI 1-minute market index and the JPY/USD 5-minute foreign exchange rate data. 
We find that the autocorrelation function from the high-frequency data exhibits a higher correlation than that from the low-frequency data.

In the next section, we describe the financial data used in the paper for analysis. In Section \ref{sec:METHODS}, 
we introduce the DFA method, the GARCH model and the AR model often used in the analysis of financial time series. 
In Section \ref{sec:RESULTS}, we present the results on the presence of the 
long-term memory property and the possible causes of its generation in the financial time series. 
Finally, we end with the summary.

\section{DATA}
We used the eight daily international market of 7 countries indices from 1991 to 2005 (from the Yahoo financial web-site) 
and the foreign exchange rates of 31 countries to the US dollar from 1990 to 2000 (from the FRB web-site), 
and as high-frequency data, the KOSPI 1-minute market index in the Korean stock market from 1995 to 2002, 
the KOSDAQ 1-minute market index from 1997 to 2004 (from the Korean Stock Exchange), and the foreign exchange rates 
of six countries to the US dollar from 1995 to 2004 (from Olson). The seven international market indices are 
USA ({S\&P} 500, NASDAQ), Hong Kong ($Hangseng$), Japan ($Nikkei 225$), Germany ($DAX$), France ($CAC 40$), UK ($FTSE 100$), 
and Korea ($KOSPI$). The daily data of the foreign exchange rates are taken from a total of 31 countries, 
including ten countries (ATS, BEF, FIM, FRF, DEM, IEP, ITL, NLG, PTE, and ESP) from 1990 to 1998, 
twenty countries (AUD, CAD, CNY, DKK, GRD, HKD, INR, JPY, KRW, MYR, NZD, NOK, SGD, ZAR, LKR, SEK, CHF, TWD, THB, and GBP) 
from 1990 to 2000, and one country (BRL) from 1995 to 2000. The 5-minute data of the foreign exchange rates were 
taken from six countries: Euro (EUR), UK (GBP), Japan (JPY), Singapore (SGD), Switzerland (CHF), and Australia (AUD).
As the financial time series data employed in this investigation, 
we used the normalized returns, $R_{t}$, from the price data, $y_{t}$, as in previous studies ;

\begin{equation}\label{e1}
{R_t} \equiv  \frac{\ln{y_{t+1}}- \ln{y_{t}}}{\sigma(r_t)},\\
\end{equation}
where $\sigma(r_t)$ is the standard deviation of the return. 
The normalized returns are composed of the magnitude time series, $|R_{t}|$, and the sign time series, $Sign_{t}$, as follows

\begin{equation}\label{e2}
{R_{k,t}}= |R_{k,t}|  \times Sign_{k,t}, \\
\end{equation}
where $R_{k,t}$ is the return series of the k-th market index calculated by the log-difference,
$|R_{k,t}|$, the magnitude series of the returns of the k-th market index, and $Sign_{k,t}$, the sign series with $+1$ for the upturn and $-1$ for the downturn.
The volatility of the returns can be studied though the magnitude series, $|R_t|$.
In this analysis, we utilized the return series and the volatility series, respectively. 
Here, for the return series, we make use of the returns, $R_{t}$, from Eq.1 and for the volatility series, 
we used its magnitude time series, $|R_{t}|$.

\section{METHODS}
\label{sec:METHODS}

\subsection{Detrended Fluctuation Analysis}
In this paper, we utilized the Detrended Fluctuation Analysis(DFA) method proposed by Peng {\em et al.} 
to quantify the long-term memory property in the financial time series [22]. 
The Hurst exponent calculated through the method of DFA can be measured as follows.
In the first step, the accumulated value after the subtraction of the mean, $\bar{x}$, from the time series, $x(i)$, is defined by

\begin{equation}\label{e3}
y(i)= \sum_{i=1}^{N} [x(i) - \bar{x}].
\end{equation}
where N is the number of the time series. In the second step, the accumulated time series is divided into boxes of the same length n. 
In each box of length n, the trend is estimated using the ordinary least square method. That is, DFA(m) is determined, where m is the filtering order. In each box, 
the ordinary least square line is expressed as $y_{n} (i)$. By subtracting $y_{n}(i)$ from the accumulated $y(i)$
in each box, the trend is removed. This process is applied to every box and the fluctuation magnitude is defined as

\begin{equation}\label{e4}
F(n)= \sqrt{\frac{1}{N} \sum_{i=1}^{N} [y(i) - y_{n}(i)]^2}.
\end{equation}

The process of the second step is repeated for every scale n and the following scaling relationship is defined by

\begin{equation}\label{e5}
F(n) \approx  cn^{H},
\end{equation}
where H is the Hurst exponent. The Hurst exponent ranges from 0 to 1, which reflects different correlation characteristics. 
If $0 \leq H < 0.5$, the time series is anti-persistent. If $0.5 < H \leq 1$, it is persistent. In the case of $H=0.5$, 
it becomes a random walk.

\subsection{Generalized Autoregressive Conditional Heteroscedasticity and Autoregressive Model}

The GARCH model proposed by Bollerslev has been widely used in the financial literature up 
to now, which includes the volatility clustering property observed in the empirical financial time series [9]. The volatility is a parameter basically used to evaluate the risk of diverse financial assets, 
so the reliable estimation of volatility is a significant task in the financial field. 
The GARCH model introduced by Bollerslev is defined as follows:

\begin{eqnarray} 
&&y_{t}= \mu + \epsilon_{t} ,~~~  \epsilon_{t} \equiv \eta_{t}\sigma_{t}, \nonumber \\ 
&&\sigma_{t}^{2} = \alpha_{0} + \sum_{i=1}^{q} \alpha_{i}\epsilon_{t-i}^{2}+ \sum_{j=1}^{p} \beta_{j}\sigma_{t-j}^{2},\label{e6} \\
&&\alpha_{0}>0, ~ \alpha_{i}, \beta_{i} ~ \geq , ~ \sum_{i} \alpha_{i}+ \sum_{j} \beta_{j} < 1, \nonumber 
\end{eqnarray}
where $\epsilon_{t}$ is a random process with zero mean, and unit variance, $y_{t}$ is the return and $\sigma_{t}^{2}$ is the volatility at time t.  As shown in Eq. 6, 
the conditional heteroscedasticity based on past volatility, $\sigma_{t}^{2}$, can be explained not only by 
the square of an error term with a lag ($\epsilon_{t-i}^{2}, i = 1,2,....,p$) but also by the conditional heteroscedasticity with 
a lag ($\sigma_{t-j}^{2}, j = 1,2,....,q$ ). In general, the case with p=1 and q=1 are often considered in empirical investigations of the financial field, 
so that the GARCH(1,1) model is used in our analysis. In addition, we obtained the standardized residual term ($\epsilon_{t}^{*} = \epsilon_{t}/ \sigma_{t}^{*}$, 
where $\sigma_{t}^{*}$ is the normalized standard deviation) after filtering the return data through the GARCH(1,1) 
model and computed the Hurst exponent by the DFA method. As the GARCH model reflects the volatility 
clustering property included in the financial time series, the residual term, $\epsilon_{t}^{*}$, turns out to be 
the time series data without the volatility clustering property. 
On the other hand, the AR(m) model reflects the autocorrelation in financial time series, which corresponds to the short-term memory.
The AR(m) model is defined by
\begin{equation}\label{e7}
r_{t} = \alpha_{0}+ \sum_{k=1}^{m}\beta_{k}r_{t-k} + \epsilon_{t},
\end{equation}
where $\epsilon_{t}$ is the normal distribution with the zero mean and the variance one. 
In particular, in the case with $\beta_{1}=1$ in the AR(1) model, $r_{t}$ follows a random walk process. 
In this paper, we employ the AR(1) model often used in the empirical studies in financial fields. After filtering 
the return data through the AR(1) model, we obtain the residual term, $\epsilon_{t}$. Since the AR(1) model 
reflects the short-term memory property contained in the financial time series, the residual 
term, $\epsilon_{t}$, becomes the time series without the short-term memory property.

\section{RESULTS}
\label{sec:RESULTS}

\begin{figure}[tb]
\includegraphics[height= 8cm, width=8cm]{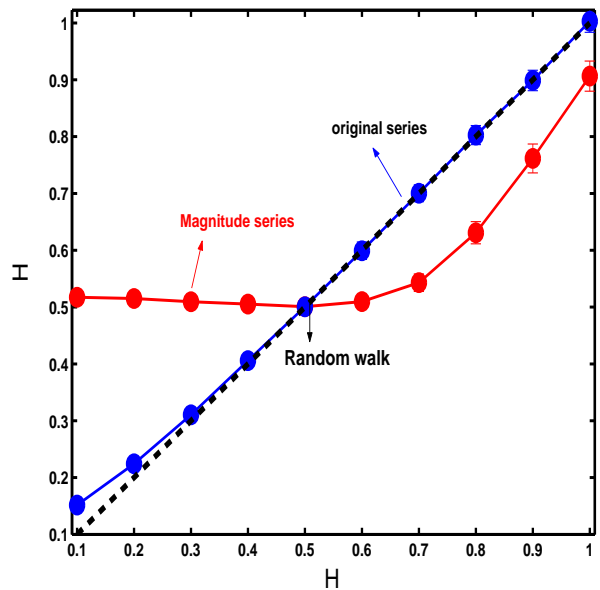}

\caption[0]{The Hurst Exponent of the original and magnitude (volatility) time series for the fractional 
brownian motion (FBM). The red circles denote the Hurst exponents of the magnitude time series, the blue circles 
the Hurst exponents of the original time series, and the dashed lines the theoretical Hurst exponents. 
Hurst exponents are averaged over 100 simulations of 20,000 data points.}
\end{figure}

\begin{figure}[tb]
\includegraphics[height=6cm, width=8cm]{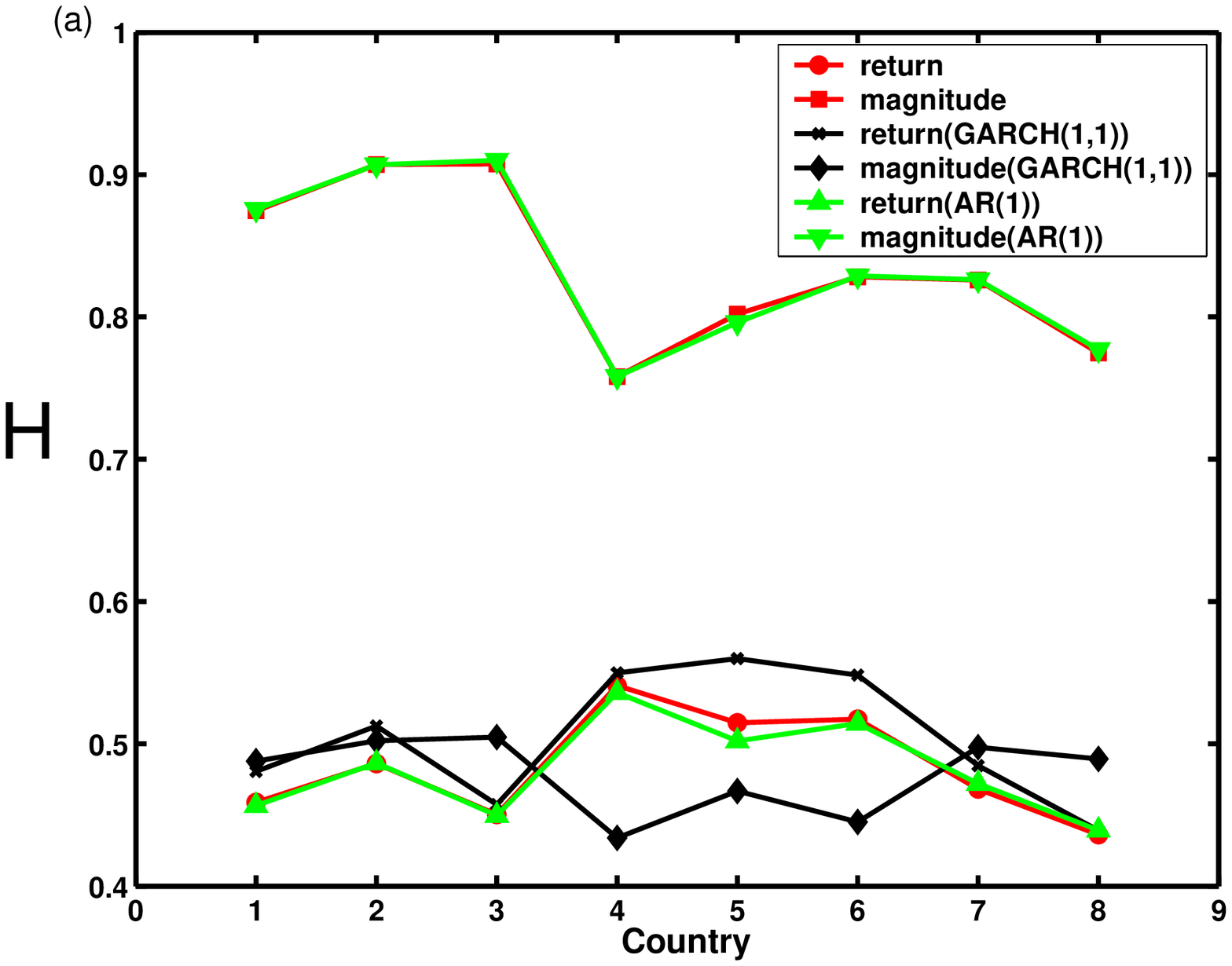}
\includegraphics[height=6cm, width=8cm]{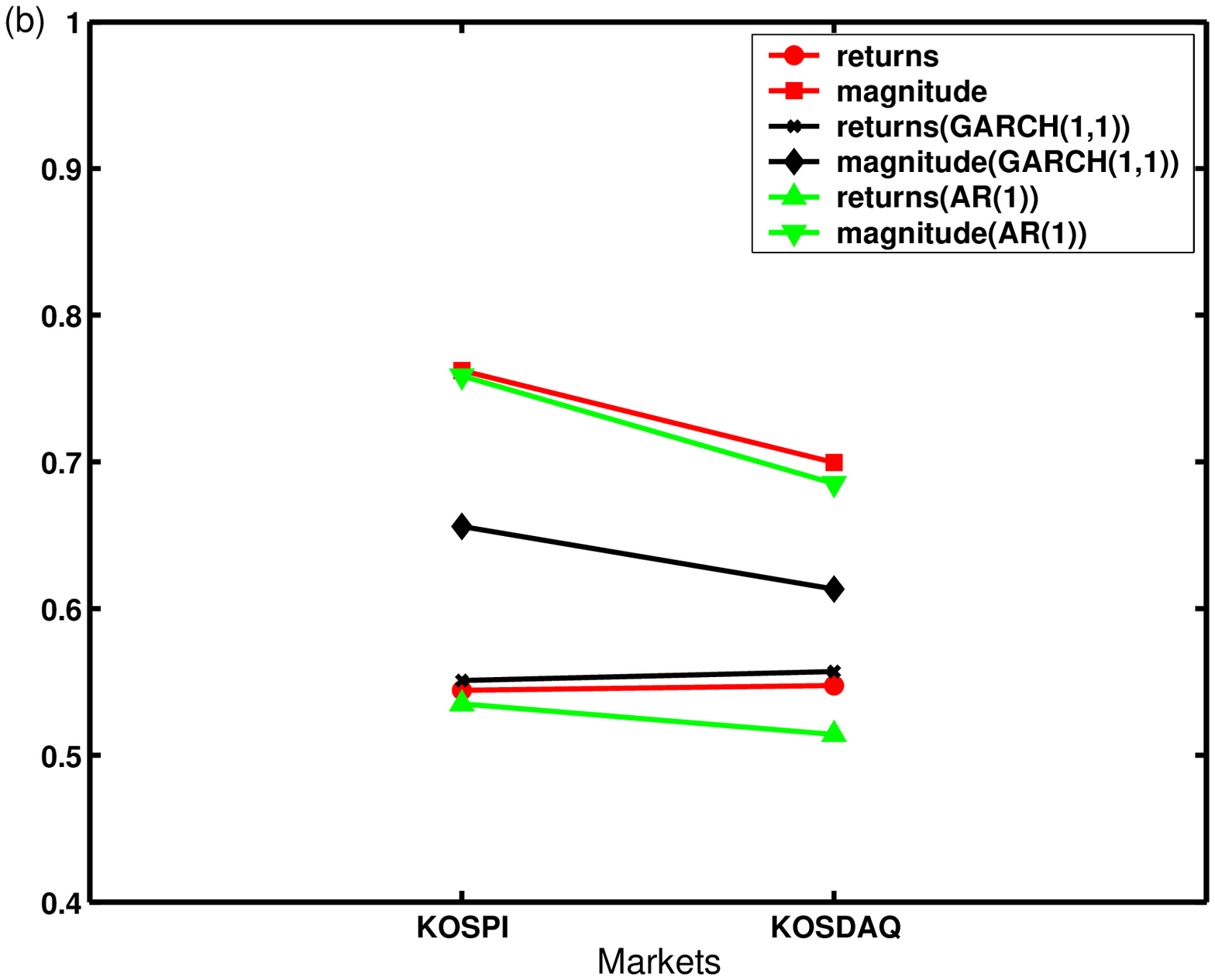}
\caption[0]{(a) Hurst exponents of the returns and volatility of international market indices 
[1: France (CAC 40), 2: Germany (DAX), 3: United Kingdom (FTSE 100), 4: Hong Kong (HangSeng), 
5: Korea (Kospi), 6: America (Nasdaq), 7: Japan (Nikkei 255), and 8: America (S\&P 500)]. 
(b) Hurst exponents of the returns and volatility of KOSPI and KOSDAQ one-minute indices. 
The red circles and squares denote the return and the magnitude of original time series. The notation (GARCH(1,1)) denotes the time series filtered by the GARCH(1,1) model(black crosses and diamonds) 
and the notation (AR(1)), by the AR(1) model.(green triangles, inverse triangles)}

\end{figure}

\begin{figure}[tb]
\includegraphics[height=6.2cm, width=8cm]{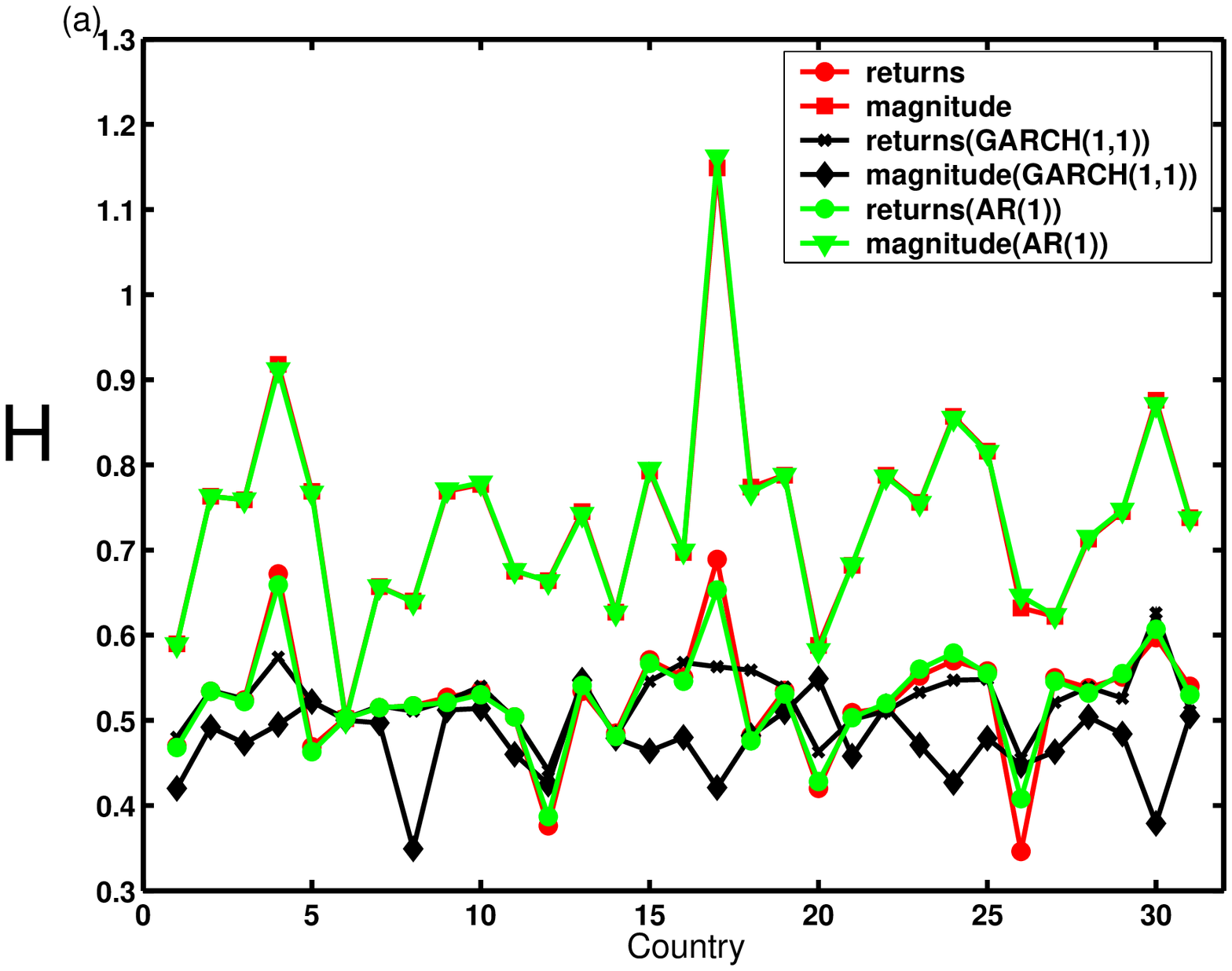}
\includegraphics[height=6cm, width=8cm]{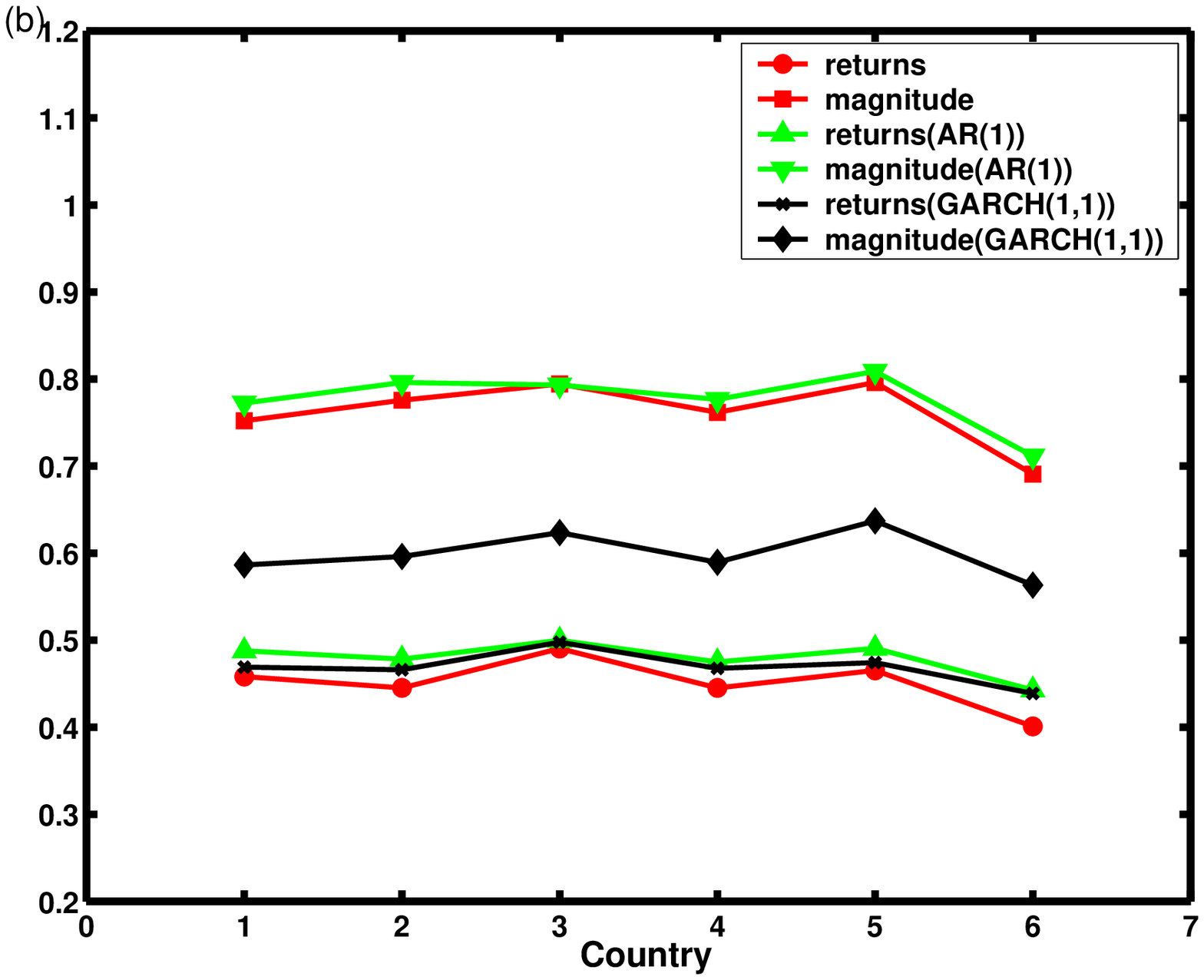}
\caption[0]{(a)Hurst exponents of the returns and volatility of daily foreign exchange rates of 31 countries. 
(b) Hurst exponents of the returns and volatility of 5-minute foreign exchange rate. 
[1:Euro(EUR), 2:UK(GBP), 3:Japan(JPY), 4:Singapore(SGD), 5:Switzerland(CHF), 6:Australia(AUD)]. 
The red circles and squares denote the return and the magnitude of original time series. The notation (GARCH(1,1)) denotes the time series filtered by the GARCH(1,1) model(black crosses and diamonds) 
and the notation (AR(1)), by the AR(1) model.(green triangles, inverse triangles)}
\end{figure}

\begin{figure}[tb]
\includegraphics[height=6cm, width=8cm]{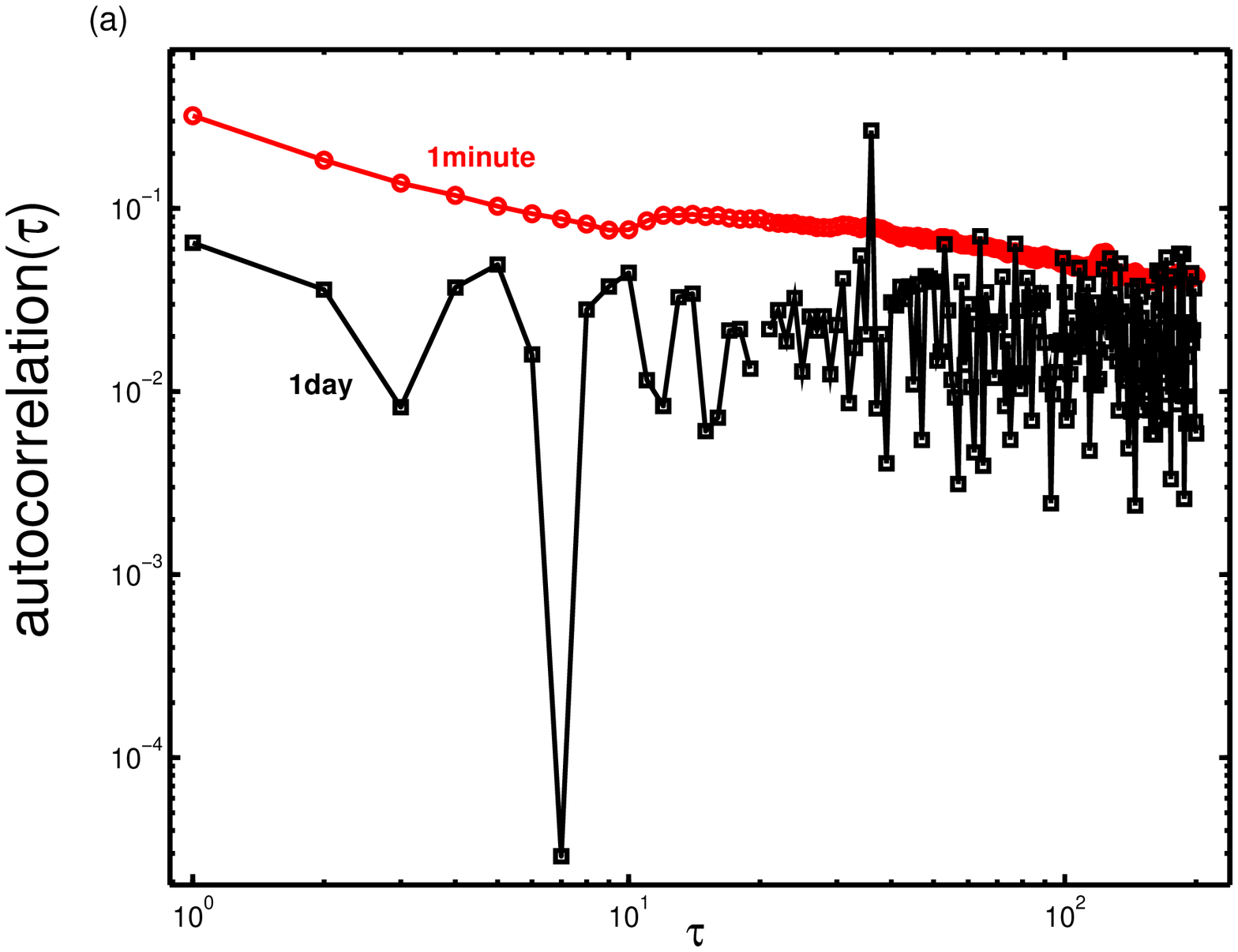}
\includegraphics[height=6cm, width=8cm]{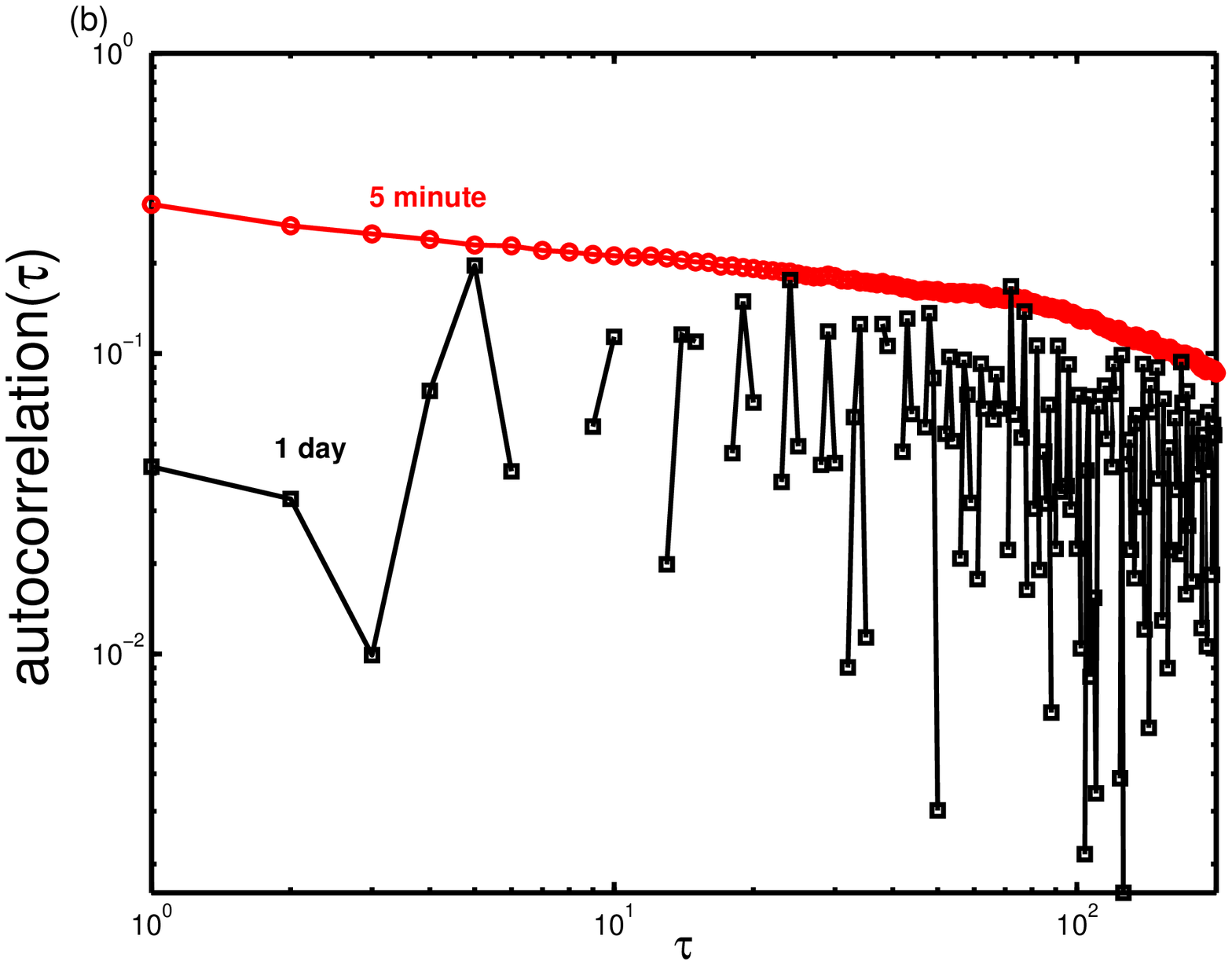}

\caption{(a) and (b) Autocorrelation functions for high-frequency and low-frequency data from 
the KOSPI 1-minute indices and JPYUSD 5-minute indices in a log-log plot. The red circles and black squares denotes high-frequency and low-frequency time series.}
\end{figure}

In this section, we investigate the long-term memory property using the DFA method in the financial 
time series of stock markets and foreign exchange markets and present the  results on the possible causes 
of its generation using data filtered through the AR(1) and GARCH(1,1) models.
First, we create time series data with only the specified memory property, with the given Hurst exponent, using the Fractional Brownian Motion (FBM) model. 
In Fig.1, we shows computed Hurst exponents through the DFA method from the data decomposed into the return time series and the volatility time series, respectively. 
In Fig.1, the Hurst exponents of the volatility time series are larger than those of the return time series 
in the region with $0 \leq H < 0.5$, but are smaller in the region with $0.5 < H \leq 1$. 
In Fig. 2, we show that the long-term memory property of the returns and the volatility by using the DFA method 
on daily and high-frequency stock market data. The  results on the possible causes of 
its generation are also presented using the data filtered through the AR(1) and GARCH(1,1) models.
Fig. 2(a) shows the Hurst exponents measured from the return and the volatility of eight daily international market 
indices. All the Hurst exponents from the original return time series (red circles), the AR(1) 
filtered data (green triangles) and the GARCH(1,1) filtered data (black crosses) give $H\simeq 0.5$, so that there is no the long-term memory property. 
On the other hand, we observe a strong long-term memory property with $0.8 \leq H \leq 0.9$
in both the original data on the volatility time series (red squares) and the AR(1) filtered data 
(green inverse triangles). Notably, we discover no the long-term memory property with $H\simeq 0.5$ in the GARCH(1,1) filtered data 
(black diamonds). These results suggest that the volatility clustering may be a possible cause for the generations of the long-term memory property 
observed in the volatility time series.

Fig. 2(b) shows the Hurst exponent measured for the returns and the volatility of the one-minute high-frequency 
market indices of Korea, KOSPI and KOSDAQ. The long-term memory property cannot be observed in the return 
time series as shown in Fig. 2(b). Furthermore, the original volatility time series (red squares) 
and the AR(1) filtered data (green inverse triangles) display a similar strong long-term memory property. 
However, in this case the GARCH(1,1) filtered data (black diamonds) still possess the long-term memory 
property with $H \approx 0.65$ albeit reduced from $H \approx 0.75$ observed in the original data. 
Note that it is hard to observe this difference in low-frequency financial time series like daily data.

Fig.3 show the results similar to ones in Fig. 2 for the case of daily and high-frequency foreign exchange market data. Fig. 3(a) shows 
the Hurst exponents from daily data for foreign exchange rates to the US dollar for 31 countries from 1990 to 2000. Fig. 3(b) shows 
the Hurst exponents from 5-minute foreign exchange rate data to the US dollar from 1995 to 2004. 
In Fig. 3, all the Hurst exponents from the original return time series (red circle), AR(1) filtered data (green triangles), 
and GARCH(1,1) filtered data (black crosses) become $H \approx 0.5$, so that the long-term memory property is not present. 
The long-term memory property, however, is found in the volatility time series (red square) and 
the AR(1) filtered data (green inverse triangles). On the other hand, in the case of the GARCH(1,1) filtered data (black diamonds), 
the long-term memory property is not observed with, $ H \approx 0.5$, in daily data. However, in the 5-minute data, 
the long-term memory property is found to persist with, $H \approx 0.6$, albeit reduced. This result is similar to the case of stock market indices.

All the financial time series data including the stock markets and the foreign exchange markets used in this investigation show 
a significant long-term memory property in the volatility time series, while not in the return time series. We found that 
the long-term memory property disappeared or diminished for the GARCH(1,1) filtered data in the volatility time series. 
This suggests that the long-term memory property in the volatility time series can be attributed to the volatility clustering empirically 
observed in the financial time series. Then, we observed the difference between the results of the GARCH(1,1) filtered data in using 
low-frequency data, like daily data, and those in using high-frequency data, like 5-minute 
or 1-minute data. This difference does not show up well in the low-frequency financial time 
series but makes us consider the possible presence of other additional factors that can be 
observed in the high-frequency financial time series. 

In Fig. 4, we shows in a log-log plot the auto-correlation function for high-frequency and low-frequency data from the KOSPI 
1-minute index from 1995 to 2002 and each of the USD/YEN 5-minute index from 1995 to 2004.
In Fig. 4 (a) and (b), the red circles and the black squares denote the high-frequency and the low-frequency data, respectively. 
We found that the autocorrelation function from high-frequency data has a higher autocorrelation than that from low-frequency data. 
The volatility clustering property is a significant factor generating the long-term memory property of the volatility time series. 
In the high-frequency financial time series, other additional attributes such as a higher correlation, besides the volatility clustering property, may be additional causes of its generation.

\section{Conclusion}

In this paper, we have investigated the long-term memory property and the possible causes of 
its generation in the return time series and the volatility time series using both low-frequency 
(daily) and high-frequency (5-minute and 1-minute) financial time series data of stock markets and 
foreign exchange markets. We employed the detrended fluctuation analysis (DFA) method to quantify 
the long-term memory property and the AR(1) and GARCH(1,1) models to remove the short-term memory 
property and the volatility clustering effect, respectively, as the possible causes of its generation.

We found that the returns time series employed in this investigation have the Hurst exponent 
with $H \approx 0.5$, while the volatility time series have a long-term memory property with Hurst exponent with $0.7 \leq H \leq 0.9$.
To investigate the possible causes of the generation of the long-term memory observed 
in this volatility time series, we employed the AR(1) and GARCH(1,1) models. We found that the 
observed long-term memory property disappeared or diminished for the GARCH(1,1) filtered data. 
This suggests that the long-term memory property observed in the volatility time series can be attributed to the volatility clustering property. 
The previous research attempted to explain qualitatively that the long-term memory property of the volatility time series 
is related to the volatility clustering effect. Our results suggest that the long-term memory of 
volatility time series is related to the volatility clustering, one of the stylized facts. 
Furthermore, in the case of the high-frequency financial time series, long-term memory 
property may be attributed to a higher autocorrelation and other factors such as long-tails in financial time series, which warrants further study.

\begin{acknowledgements}
This work was supported by a grant from the MOST/KOSEF to the National Core Research Center for Systems Bio-Dynamics (R15-2004-033), 
and by the Korea Research Foundation (KRF-2005-042-B00075), and by the Ministry of Science \& Technology through the National Research Laboratory Project, and by the Ministry of Education through the program BK 21.
\end{acknowledgements}

\end{document}